\documentstyle[aps,prl,epsfig,preprint]{revtex}
\begin{document}
\date{\today}
\title {Order in driven vortex lattices in superconducting Nb films with nanostructured pinning potentials}
\author{M. V\'elez$^{a,}$\cite{oviedo}, D. Jaque$^{a}$, 
J. I. Mart\'\i n$^{a,}$ \cite{oviedo},
F. Guinea$^{b}$, and J. L. Vicent$^{a}$}
\address{$^{a}$Depto. F\'\i sica de Materiales, F. F\'\i sicas, Universidad
Complutense, 28040 Madrid, Spain}
\address{$^{b}$Instituto de Ciencia de  Materiales
de Madrid, CSIC, Cantoblanco, 28049 Madrid, Spain }
\date{\today}
\maketitle

\tightenlines

\widetext



\begin{abstract} 
 Driven vortex lattices have been studied in a  material
with strong pinning, such as  Nb films. Samples in which 
natural random pinning coexists with artificial 
ordered arrays of defects (submicrometric Ni dots) 
have been fabricated with different geometries 
(square, triangular and rectangular). Three different 
dynamic regimes are found: for low vortex velocities, 
there is a plastic flow regime in which random defects 
frustrate the effect of the ordered array; then, 
for vortex velocities in the range 1-100 m/s, 
there is a sudden increase in the interaction between 
the vortex lattice and the ordered dot array, 
independent on the geometry. This effect is associated to the 
onset of quasi long range order in the vortex lattice 
leading to an increase in the overlap between the vortex 
lattice and the magnetic dots array. Finally, at larger velocities
the ordered array-vortex lattice interaction is suppressed again,
in agreement with the behavior found in numerical simulations.
\end{abstract}

\narrowtext
\tightenlines

The influence of both applied magnetic field and driving current 
on the type II superconductors behavior 
is of great scientific and technological interest.  
When the vortex lattice is driven by an applied 
current in the presence of defects, 
it can adopt several dynamic flow regimes depending 
on the driving current and therefore on the vortex velocity. 
Transport measurements \cite{transport}, 
neutron scattering \cite{neutron} and Bitter decoration 
experiments \cite{bitter1,bitter2} 
have provided strong evidence of these different dynamics phases,
including creep, plastic flow and ordered (elastic) flow, 
and even of the existence of several phase 
transitions as a function of vortex 
velocity \cite{teoria1,teoria2}. Due to this complex behavior, 
the vortex lattice emerges as an excellent model system for 
many interesting fields in condensed matter physics 
such as 
self-organized criticality, atomic scale friction and phase transitions 
in the presence of disorder/order\cite{crit,kawaguchi}.

Theoretical works on driven vortex lattices, 
in the presence of either random \cite{random} or 
ordered \cite{ordered1,ordered2} pinning centers, have 
predicted rich phase diagrams as a function of vortex 
velocity and pinning strength. In general, most of the 
experimental studies have focused on dynamic phases in the 
presence of weak random pinning centers in materials 
such as 2$H$-NbSe$_2$ single crystals \cite{nbse} or 
amorphous Nb$_3$Ge films \cite{nbge}  where the low values 
of the critical current allow experimental access to a 
wide portion of the relevant force-velocity space. 
However, driven vortex lattices materials with strong pinning,
such as Nb, have received much less attention. 
In these systems, the fabrication of ordered arrays 
of artificial pinning centers by submicrometric 
lithographic techniques have proved to be a very useful 
tool in the study of vortex dynamics under 
tailored nanostructured pinning potentials \cite{bruyn,martin1,ketterson,martin2,metlushko}. 

In particular, depending on the size and material 
of the artificial pinning centers, the relative strength 
of the ordered and disordered pinning potentials 
in the sample can be tuned in a controlled way \cite{axel}.

	In this work, we present results obtained from transport 
measurements on  Nb films with ordered arrays of 
magnetic dots in order to study the dynamic interactions 
of the vortex lattice with random and ordered defects. 
Deformations in the vortex lattice frustrate the interactions 
with the artificial dot array for slow vortex motion, 
until the onset of ordering in the vortex lattice 
takes place for vortex  velocities in the 1-10 m/s range.  
The evidence of this transition  has been found to be 
independent on array geometry, indicating that the ordered 
array of dots is acting as a probe of the intrinsic behavior 
of the driven vortex lattice. Finally, for velocities above 100 m/s
the effect of the ordered array on the moving vortex lattice
disappears, in good agreement with numerical simulations.

Magnetic (Ni) dots approximately 200 nm in diameter 
and 40 nm in thickness were prepared on Si(100) 
substrates using electron beam lithography combined 
with a lift-off process described 
elsewhere \cite{martin1,jmmm}. After this, 
a 100 nm thick Nb film was sputtered on top. 
Finally optical lithography and reactive ion 
etching were used to define a 80 $\mu$m wide bridge 
for transport measurements.  Three different 
array geometries were fabricated:  $0.45 \times 0.45 \mu$m$^{2}$ 
square array, $0.35 \times 0.5 \mu$m$^{2}$ rectangular array 
and a triangular array with a lattice constant of 0.4 $\mu$m. 

In the case of the rectangular array, transport measurements 
were carried out by applying current parallel to the shortest 
lattice dimension. The dc measurements were performed 
in a helium cryostat with a 9T superconducting magnet. The 
magnetic field is always applied along the film normal.

\begin{figure}[h]
\epsfxsize=12cm
\centerline{\epsffile{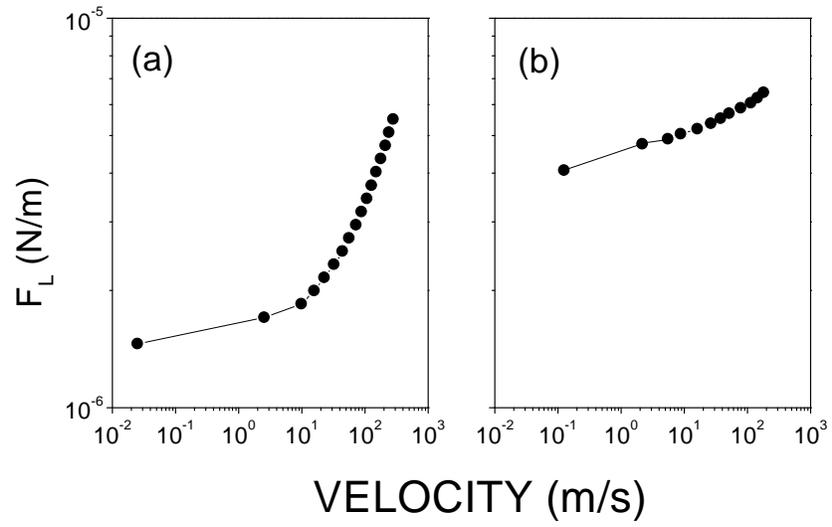}}
\caption{M. Velez et al. }
\label{fig_1}
\end{figure}

As a first approach to the dynamics of driven vortex lattices 
we can consider that, in the presence of a driving current, 
the vortex lattice is set into motion with average velocity 
{\bf v} by the Lorentz force ${\bf F_L} = \Phi_0 {\bf J} \times {\bf n}$, 
where {\bf J} is the transport current, $\Phi_0$ is 
the quantum of flux, and {\bf n} is the unit vector 
along the field direction. The vortex velocity {\bf v} 
is determined  by the balance between the viscous, 
pinning and Lorentz forces as \cite{vinokur} 
\begin{equation}
\eta {\bf v} +\alpha {\bf v}\times{\bf n} = 
{\bf F_L} + \langle {\bf F_{P}} \rangle  \label{balance}%
\end{equation}
where $\eta$ is the Bardeen-Stephen viscosity, 
$\alpha$ is the viscous Hall coefficient,  
and $\langle {\bf F_P} \rangle$ is the pinning force averaged 
over vortex positions in time.
In the dirty limit, $\alpha << \eta$ and the viscous Hall 
force can be neglected. The resulting 
force-velocity curves ($F_L$ $vs$ $v$) can be simply 
derived from experimental I-V characteristics calculating the Lorentz force and using the Josephson relation for the electric field 
${\bf E}= {\bf B} \times {\bf v}$, as shown in Fig. 1 for a Nb 
film with and without periodic pinning centers. For 
low velocities the curves approach a constant force value 
that corresponds to the static pinning force $F_{P0}$, 
related to the critical current $J_C$ by $F_{P0}=\Phi_0 J_C$. 
The free flux flow behavior, characterized by a linear dependence 
of $F_L$ on $v$ should be recovered for very high velocities, 
where the interaction between the pinning centers and the vortex 
lattice becomes negligible. However pure free flux flow is usually 
only accessible experimentally by fast ramp current methods, 
even in relatively weak pinning materials \cite{Andrei}. 
Often, in a wide range of vortex velocities 
such as plotted in Fig.\ref{fig_1}, 
the force velocity dependence is found to be much weaker than linear, 
revealing that the term corresponding to pinning interactions 
$\langle {\bf F_P} \rangle$ is of the same order as the Bardeen-Stephen viscous force.

 The calculation of  $\langle {\bf F_P} \rangle$ is not 
straightforward, since interactions among the vortices and 
with the pinning centers induce dynamic deformations in the 
vortex lattice that strongly affect this average. The orientation
of $\langle {\bf F_P} \rangle$ 
is opposite to the 
vortex velocity \cite{vinokur}, i.e. it can be written as 
$\langle {\bf F_P} \rangle=-\gamma(v) {\bf \hat{v}}$, 
with $\gamma(v)$ a velocity dependent coefficient 
and ${\bf \hat{v}}$ the unit vector along {\bf v}. At sufficently
large driving currents, the time average of the pinning force
is related to the spatial average by $\langle {\bf F_P} \rangle_{\rm time}
\approx \langle | {\bf F_P}^2 | \rangle_{\rm position} / | {\bf v} |$. The equation of motion of the vortex lattice is given by
\begin{equation}
\eta {\bf v} +\gamma(v) {\bf \hat{v}} = {\bf F_L} \label{friction}%
\end{equation}
Equation (\ref{friction}) implies that the effect of pinning 
centers can be assimilated to a kinetic viscous force, 
in a similar way to the problem of friction between 
two solids \cite{kawaguchi} (in this case, the vortex 
lattice and the pinning array).

 The distribution and strength of the pinning centers  
affect both the magnitude and velocity dependence of 
$\langle {\bf F_P} \rangle$. This is clearly observed, 
for example, in differences between the two $F_L-v$ 
curves in Fig.\ref{fig_1}, that were measured in two adjacent 
regions of the same Nb film, with and without periodic pinning centers.
Further insight in the behavior of the vortex lattice 
can be obtained by focusing on the differences 
in $\langle {\bf F_P} \rangle$ depending on the 

commensurability between the vortex lattice and periodic array of dots.

\begin{figure}[h]
\epsfxsize=20cm
\centerline{\epsffile{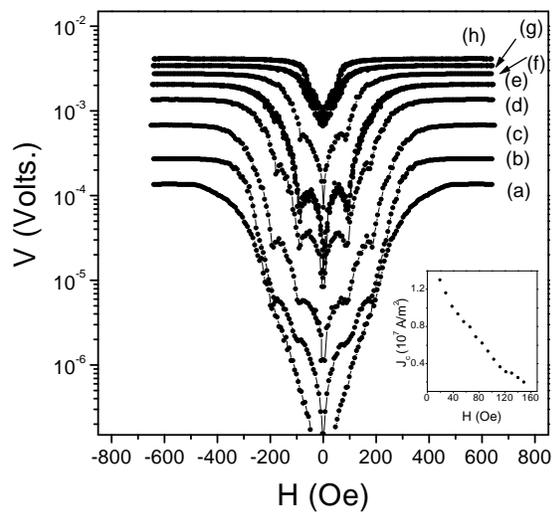}}
\caption{M. Velez et al.}
\label{fig_2}
\end{figure}

A general view of the effect of commensurability 
is shown in Fig.\ref{fig_2}, where we have plotted the 
voltage versus magnetic field curves obtained for different 
current densities in the range $10^7 - 10^9$ A/m$^2$ 
for the Nb film with the square array of magnetic dots. 
Magnetoresistence minima are observed  at equal
magnetic field intervals, $\Delta H_m$=98 Oe. 
These can be explained \cite{martin1} by a geometric matching 
between the vortex lattice and the magnetic dot array. 
It should be noted that the current dependence of the 
magnetoresistance is not the same whether the measuring field 
corresponds to a matching condition or not. These differences show up in the fact that the magnetoresistance minima are much deeper in an intermediate current range than at low and high currents. Actually,  
the critical current, shown in the  inset of Fig.\ref{fig_2}, presents a monotonous 
decrease as a function of field, without any special feature at 
the matching field values. These magnetoresistance and critical 
current data are clearly indicating the existence of several different regimes: first, at $v=0$, the static pinning force $F_{P0}$ is dominated 
by the random pinning centers, that frustrate the effect of the 
ordered array of dots; then, in the driven vortex lattice the 
kinetic viscous force due to the ordered array of pinning centers 
$\langle {\bf F_P} \rangle$ is enhanced in a certain velocity range, and disappears at large driving currents. 
This behavior is characteristic of small dot diameter arrays 
where periodic pinning is weaker than random pinning, 
and it is also observed for the rectangular and triangular arrays 
used in this study, but with $\Delta H_m$ = 120 and 145 Oe, 
respectively, in good agreement with theoretical predictions of the matching conditions.

\begin{figure}[h]
\epsfxsize=12cm
\centerline{\epsffile{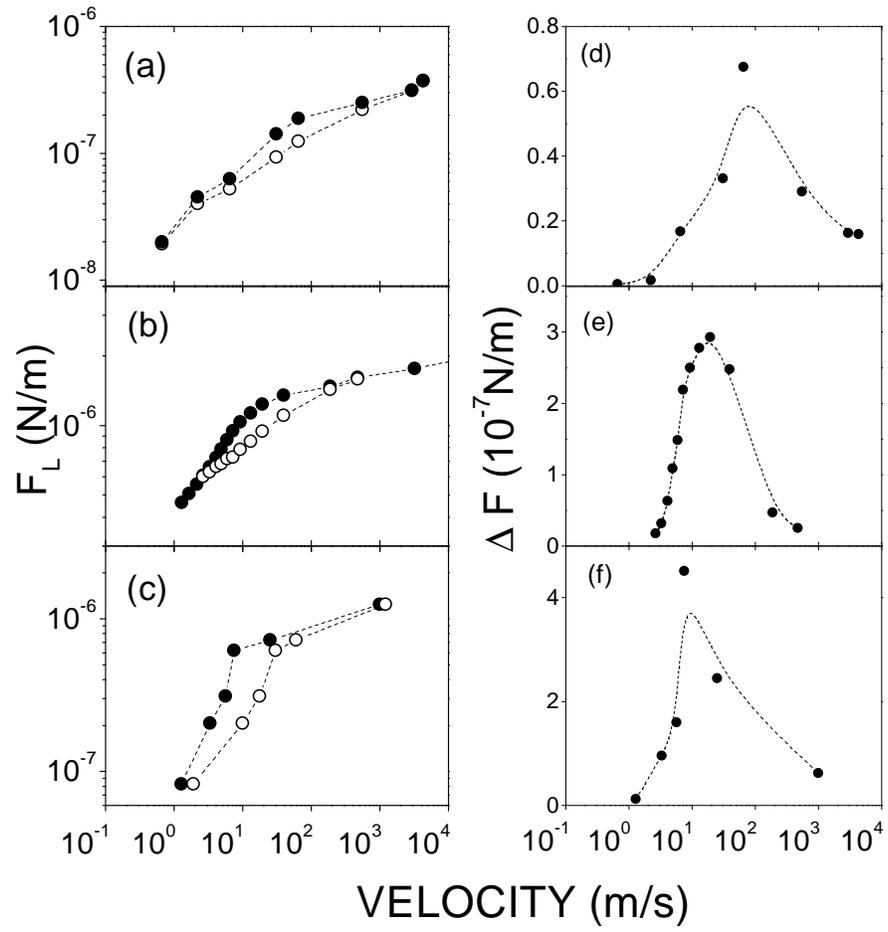}}
\caption{M. Velez et al.}
\label{fig_3}
\end{figure}

	Figures \ref{fig_3}(a)-(c) show the force-velocity curves in 
the driving current range where commesurability minima are observed, 
derived from the I-V characteristics, for the Nb films with square, 
triangular and rectangular  arrays of dots measured at 
two different fields: at the first matching field $H_{comm}$ 
for each array (filled symbols) and at a smaller field 
value $H_{incomm}$ close to $H_{comm}$  but far enough to be 
out of matching conditions (hollow symbols). The obtained curves 
present quite similar values for low and high vortex velocities. 
At intermediate vortex velocities (1-100 m/s) the matching curve 
differs from that obtained at no matching fields. In all cases, 
the overall dependence is clearly weaker than linear, 
indicating that the vortex lattice is far from the free flux 
flow regime. Two kinds of force contributions can be considered 
in this system; the sum of the Bardeen-Stephen viscous force 
and the interaction with random defects 
($\eta v + \langle F_{P,random} \rangle$), and the average 
interaction force from the periodic array 
$\langle F_{P,ordered} \rangle$, which is a strongly peaked 
function of the magnetic field at the matching condition. 
In a first approximation, taking into account that in 
these samples random pinning is much stronger than periodic 
pinning, $\langle F_{P,ordered} \rangle$ can be assumed to be 
negligible for the incommensurate vortex lattice; 
then  the difference 
between the filled and hollow symbols curves can be used to extract the 
value of $\langle F_{P,ordered} \rangle$ at the matching conditions.

 This force increment, 
$\Delta F = F(H_{comm})-F(H_{incomm})$ = 
$\langle F_{P,ordered} \rangle$ is shown in Figs. 3(d)-(f) for the three arrays considered in this 
work. There is a clear onset of the interactions with 
the ordered pinning array in the range 1-10 m/s, 
observable in the three cases. This onset is almost independent 
on the ordered array geometry, so that it must be related 
with an intrinsic change in the properties of the driven 
vortex lattice. The increase in $\langle F_{P,ordered} \rangle$ can 
be directly correlated with an enhancement in the long range order 
in the vortex lattice, since this force should be zero for a 
completely disordered vortex state. Therefore, the data in 
Fig. 3 show the crossover between two different regimes: 
first, for slow vortex motion $\langle F_{P,ordered} \rangle = 0$, 
implying a plastic motion regime where the effect of the ordered 
array is frustrated by random pinning and, then, 
for $v \sim$ 10 m/s the lattice flows in a more ordered state 
in which it has a better overlap with the magnetic dot array.

In general, theoretical simulations of driven vortex 
lattices have predicted \cite{random,ordered1}  two kinds of 
ordered states: a smectic glass phase, with quasi long range 
order only in the direction transverse to the motion, and a 
Bragg glass phase at higher velocities that is also ordered along 
the longitudinal direction. It is worth to note that 
the ordered pinning effect observed here occurs at $H_{comm}$ 
corresponding to matching between the vortex lattice 
with the two dimensional dot array cell (this is particularly 
evident in the case of the rectangular array, where 
$H_{comm}= 120 $ Oe corresponds to a vortex density 
$5.8 \times 10^8$  cm$^{-2}$ in good agreement with the 
density of pinning centers $(0.35 \times 0.5 \mu$m$^2)^{-1}$= 
5.7 $\times 10^8$ cm$^{-2}$). This implies that both the 
longitudinal and transverse correlation lengths in the vortex 
lattice are at least as large as the array cell dimension when 
the peak in $\Delta F$ appears, suggesting the 
presence of a Bragg glass phase. The ordering velocity that 
we have found in these Nb films is two orders of magnitude 
higher than those obtained previously for other superconducting 
systems \cite{Andrei} such as $2H$-NbSe$_2$ 
(typically of the order of 0.1 m/s). These differences can 
be correlated with the higher values of the critical current 
in our samples that imply stronger random pinning potentials 
($J_C \approx 10^7$ A/m$^2$ in the Nb films presented in 
this work, while $J_C \approx 10^5$ A/m$^2$ in \cite{Andrei}).

\begin{figure}[h]
\epsfxsize=12cm
\centerline{\epsffile{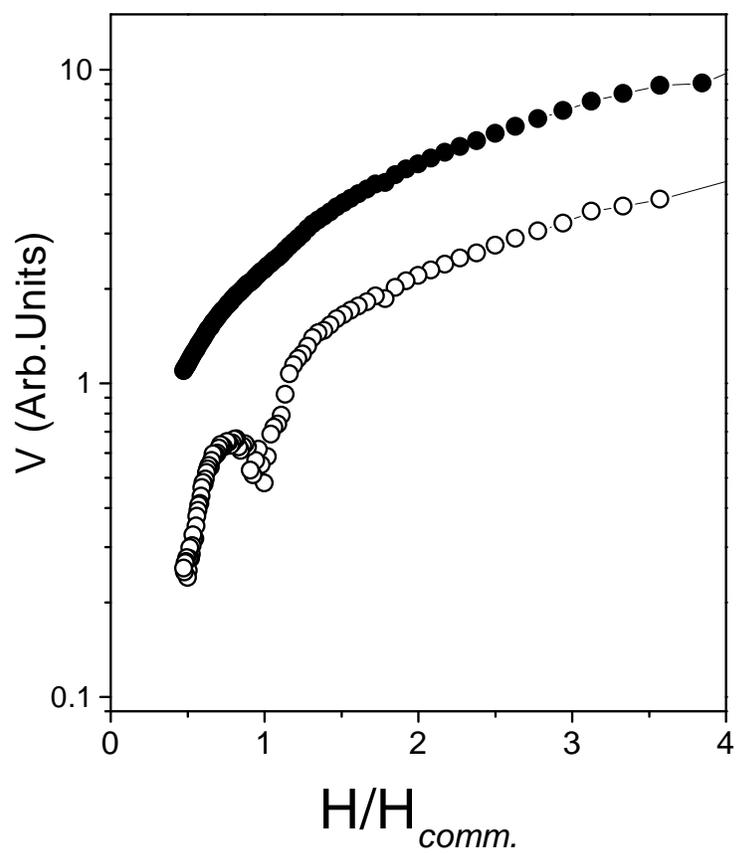}}
\caption{M. Velez et al.}
\label{fig_5}
\end{figure}

 Finally, it has to be noted that for vortex velocities 
above 100 m/s, $\Delta F$ decreases again, i.e. the interaction between the
ordered array and the vortex lattice is suppressed. We have tried to understand this large velocity regime  by modelling the moving
vortex lattice by a 1D array of point particles
which move under a driving force in a combination of
a periodic potential and random pinning centers.
A similar setup, in 2D, was
considered in\cite{kawaguchi}, although the issue of
commensurability was not addressed.  Our model is an approximation to
the 1D chains with short range order which are expected in
a driven smectic\cite{BMR98}. The results are shown in Fig.\ref{fig_5}.
Commensurability effects appear only at one matching field, due
probably to the length of the chain used (20-30 vortices).
It is interesting to note that, as in the experimental data shown in Fig.2, these
effects dissappear at large driving currents.
The simulations show that there are, at least, three
different regimes: i) At low driving currents, the lattice
remains disordered at all fields, leading to 
structureless V-B curves. ii) At intermediate currents,
the vortex lattice shows some degree of order, which increases
as a function of the field. In this regime the minimum in the V-B curves at the matching field is most pronounced. iii) Finally, at high driving currents, the lattice is very well ordered, with no substantial dependence on the applied field. In this regime, the effects at H$_{\rm comm}$ are very weak, if any.  It is worth to note that the same  effect of reduction in the interaction force at high velocities has been numerically predicted in simulations of the similar problem of friction between two lattices at atomic scale \cite{kawaguchi}.

	In summary, we have analyzed the force-velocity 
characteristics for several Nb films with square, triangular 
and rectangular arrays of Ni dots, in which ordered and disordered 
defects coexist. By comparison of the  curves at matching 
magnetic field and away from matching conditions we have 
found that there is a clear increase in the interactions 
between the vortex lattice and the ordered array of dots 
in the vortex velocity range 1-100 m/s. The sudden 
enhancement in $\langle F_{P,ordered} \rangle$  has 
been attributed to the onset of quasi long range order 
in the vortex lattice  at this driving velocity. 
This result is independent on the array geometry, 
indicating that a change in the intrinsic 
properties of the vortex lattice is being observed. For very high vortex velocities,
commensurability effects are suppresed again, in agreement with numerical
simulations.    

\acknowledgments

     This work has been supported by the 
Spanish CICYT (grant MAT99/0724, and PB96/0875) and by the European Science Foundation VORTEX program. We thank G. W. Crabtree for useful comments.

\newpage

FIG. 1. Force vs velocity curves derived from the I-V 
characteristics for (a) a simple Nb film and (b) a Nb 
film on a rectangular array of Ni dots, grown on the 
same substrate, at $H$= 81 Oe and $T=0.98T_C$. 
This field corresponds to the first matching 
field of the rectangular array.
\\

FIG. 2. (a) $V-B$ curves obtained on the square array at 
$T = 0.99 T_C$ for different current densities ($J = 6.25 
\times 10^6,$$ 1.25 \times 10^7, 3.1 \times 10^7,$ 
$6.25 \times 10^7, 9.4 \times 10^7,$ $1.25 \times 10^8, 
1.56 \times 10^8$ and $1.87 \times 10^9 A/m^2$ for the 
(a)-(h) curves respectively). Inset shows the field dependence of
the critical current at the same temperature.
\\

FIG. 3. (a) Force-velocity characteristics  for (a) a 
Nb film on a $0.45 \times 0.45 \mu$m$^{2}$ square 
array of dots (filled symbols, $H_{comm}$ = 98 Oe; 
hollow symbols, $H_{incomm}$ = 60 Oe) (b) a Nb film on a 
triangular array of dots of lattice constant $0.4 \mu$m 
(filled symbols, $H_{comm}$ = 145 Oe; hollow symbols, 
$H_{incomm}$ = 114 Oe) (c) a Nb film on a $0.35 \times 0.5 \mu$m$^{2}$ 
rectangular array of dots (filled symbols, 
$H_{comm}$ = 120 Oe; hollow symbols, $H_{incomm}$ = 90 Oe); (d), (e), (f) Force increment $\Delta F = F(H_{comm}) - F(H_{incomm})$ 
(i.e. from no matching conditions to matching conditions) 
as a function of vortex velocity for the Nb films in (a), (b) and (c) respectively. Lines are guides to the eye.
\\

FIG. 4. $V-B$ characteristics of a model of moving vortices
under the influence of a periodic potential and random pinning
(see text), corresponding to: filled symbols,  $J = 10 J_0$ and hollow symbols, $J = 2.5 J_0$, where $J_0$ is the critical current at 0.6 $H_{comm}$.

\begin{references}
\bibitem[*]{oviedo} Present address: Depto. F{\'\i}sica, 
Universidad de Oviedo, Spain.
\bibitem{transport} S. Battacharya and M. J. Higgins, 
Phys. Rev. Lett. {\bf 70} 2617 (1993).
\bibitem{neutron}U. Yaron, P. L. Gammel, D. A. Huse, R. N. Kleiman, 
C. S. Oglesby, E. Bucher, B. Batlogg, D. J. Bishop, K. Mortensen, K. Clausen, C. A. Bolle and F. de la Cruz, Phys. Rev. Lett. {\bf 20}, 748 (1994)
\bibitem{bitter1}  F. Pardo, F. de la Cruz, P. L. Gammel, 
E. Bucher and D. J. Bishop, Nature {\bf 396}, 6709 (1998). 
\bibitem{bitter2} M. Marchevsky, J. Aarts and P. H. Kes, 
Phys. Rev. B {\bf 60}, 14601 (1999). 
\bibitem{teoria1} A. E. Koshelev and V. M. Vinokur, Phys. Rev. Lett. 
{\bf 73}, 3580 (1994)
\bibitem{teoria2} L. Balents and M. P. A. Fisher, 
Phys. Rev. Lett. {\bf 75}, 4270 (1995).
\bibitem{crit}K. E. Bassler and M. Paczuski, Phys. Rev. Lett. {\bf 81}, 3761 (1998).
\bibitem{kawaguchi} T. Kawaguchi and H. Matsukawa, 
Phys. Rev. B {\bf 61}, R16366 (2000).
\bibitem{random}P. L. Doussal and T. Giamarchi, 
Phys. Rev. B {\bf 57}, 11356 (1998).
\bibitem{ordered1} C. Reichhardt, C. J. Olson, and 
F. Nori, Phys. Rev. B {\bf 58}, 6534 (1998);  
C. Reichhardt and G. T. Zimanyi, Phys. Rev. B {\bf 61}, 14354 (2000).
\bibitem{ordered2}V. M. Marconi and D. Dominguez, 
Phys. Rev. Lett. {\bf 82}, 4922 (1999).
\bibitem{nbse} W. Henderson, E. Y. Andrei, 
M. J. Higgins and S. Battacharya, Phys. Rev. Lett. {\bf 80}, 381 (1998).
\bibitem{nbge} P. Berghuis, A. L. F. 
van der Slot and P. H. Kes, Phys. Rev. Lett. {\bf 65}, 2583 (1990).
\bibitem{bruyn} M. Baert, V. V. Metlushko, 
R. Jonckheere, V. V. Moshchalkov and Y. Bruynseraede, 
Phys. Rev. Lett. {\bf 74}, 3279 (1995).
\bibitem{martin1} J. I. Mart\'\i n, M. V\'elez, 
J. Nogu\'es, and I. K. Schuller, Phys. Rev. Lett. {\bf 79}, 1929 (1997).
\bibitem{ketterson} D. J. Morgan and J. B. Ketterson, 
Phys. Rev. Lett. {\bf 80}, 3614 (1998).
\bibitem{martin2} J. I. Mart\'\i n, M. V\'elez, 
A. Hoffmann, I. K. Schuller and J. L. Vicent, 
Phys. Rev. Lett. {\bf 83}, 1022 (1999).
\bibitem{metlushko} V. Metlushko, U. Welp, G. W. Crabtree, 
R. Osgood, S. D. Bader, L. E. DeLong, Z. Zhang, 
S. R. J. Brueck, B. Ilic, K. Chung, and 
P. J. Hesketh, Phys. Rev. B. {\bf 60}, R12585 (1999).
\bibitem{axel} A. Hoffmann, P. Prieto and I. K. Schuller, 
Phys. Rev. B {\bf 61}, 6958 (2000).
\bibitem{jmmm} J. I. Mart\'\i n, Y. Jaccard, A. Hoffmann, J. Nogu\'es, J. M. George, J. L. Vicent, 
and I. K. Schuller, J. Appl. Phys. {\bf 84}, 411 (1998).
\bibitem{vinokur}V. M. Vinokur, V. B. Geshkenbein, 
M. V. Feigel'man and G. Blatter, Phys. Rev. Lett. {\bf 71}, 1242 (1993).
\bibitem{Andrei} Z. L. Xiao, E. Y. Andrei, 
P. Shuk and M. Greenblatt, Phys. Rev. Lett. {\bf 85}, 3265 (2000).
\bibitem{BMR98}
L. Balents, M. C. Marchetti and L. Radzihovsky, Phys. Rev. B
{\bf 57}, 7705 (1998).
\end{references}
\end{document}